\documentclass[12pt]{article}

\usepackage{slashed}
\usepackage{times}
\usepackage{graphics}

\usepackage{oldlfont}

\declareslashed{}{/}{.1}{0}{E}

\def\sss{\scriptscriptstyle}

\def\N{{\cal N}}

\def\E{{\cal E}}
\def\B{{\cal B}}

\def\psib{\overline\psi\hspace{-2.6mm}\phantom{\psi}}
\def\k{\kappa}

\def\D{{\cal D}}

\def\r{\rho}
\def\s{\sigma}

\def\Sl#1{\slashed{#1}}

\def\wt#1{\widetilde{#1}}

\def\ss{\scriptscriptstyle}

\def\d{\partial}
\def\m{\mu}
\def\n{\nu}

\def\e{\epsilon}

\def\F{{\cal F}}

\def\be{\begin{equation}}
\def\ee{\end{equation}}
\def\beq{\begin{equation}}
\def\eeq{\end{equation}}
\def\bea{\begin{eqnarray}}
\def\eea{\end{eqnarray}} 
\def\beqa{\begin{equation}\begin{array}{l}}
\def\eeqa{\end{array}\end{equation}}

\def\eqn#1{(\ref{#1})}

\def\eqref#1{eq.~(\ref{eq:#1})}

 \def\G{{\it\Gamma}} \def\g{\gamma}

\def\L{{\it\Lambda}}

\def\w{\omega}

\def\nn{\nonumber}

\newcommand{\DL}[1]{\Delta^{\!(#1)}}

\begin{document}

\thispagestyle{empty}
\begin{flushright}
\framebox{\small BRX-TH~478
}\\
\end{flushright}

\vspace{.8cm}
\setcounter{footnote}{0}
\begin{center}
{\Large{\bf 
Inconsistencies of Massive Charged Gravitating Higher Spins}
    }\\[10mm]

{\sc S. Deser$^\sharp$
and A. Waldron$^\flat$
\\[6mm]}

{\em\small  
${}^\sharp$Physics Department, Brandeis University,\\ 
Waltham, MA 02454, 
USA\\ {\tt deser@brandeis.edu}}\\[5mm]

{\em\small ${}^\flat$Department of Mathematics, University
of California,\\
Davis, CA 95616, USA\\
{\tt wally@math.ucdavis.edu}}\\[5mm]

{\small (\today)}\\[1cm]

{\sc Abstract}\\
\end{center}

{\small
\begin{quote}

We examine the causality and degrees of freedom (DoF) problems
encountered by charged,
gravitating, massive higher spin fields.
For spin $s$=3/2, making the metric dynamical  yields improved
causality bounds. These involve only the mass, the product $eM_P$ of
the charge and Planck mass and the cosmological
constant $\Lambda$. The bounds are themselves related to a 
gauge invariance of the timelike component of the
field equation at the onset of acausality. While
propagation is causal in arbitrary E/M backgrounds, the allowed
mass ranges of parameters are of Planck order.
Generically, interacting spins $s>3/2$ are subject to 
DoF violations as well as to acausality; the former
must be overcome before analysis of the latter can even
begin.  Here we review both difficulties
for charged $s=2$ and show that while a $g$-factor of $1/2$
solves the DoF problem, acausality persists for any $g$.
Separately we establish that no $s=2$ theory --
DoF preserving or otherwise -- can be tree unitary.

\bigskip

\bigskip


\end{quote}
}

\newpage





\section{Introduction}

Localized, massive $s>1$ particles have never been observed,
in agreement with a 
large (if somewhat confusing) higher
spin lore that they cannot
be made to interact consistently even with
gravity or electromagnetism.
Here we combine two earlier lines of analysis
into a systematic study of higher spins
coupled to Einstein--Maxwell fields.

A previous examinations of neutral higher spins
propagating in cosmological, as well as $s=3/2$ in 
electromagnetic (E/M), 
backgrounds
reveal the following:
(i) Massive higher spins propagate
consistently in constant curvature backgrounds for a range of
parameters $(m^2,\Lambda)$ centered around the Minkowski line
$(m^2,0)$~\cite{Deser:2001xr,Deser:2001wx,Deser:2001us,
Deser:2001pe,Deser:2000de}. 
(ii) The original unitarity~\cite{Johnson:1961vt} 
(or equivalently causality~\cite{Velo:1969bt})
difficulties of massive $s=3/2$ persist 
in pure E/M backgrounds,
even including all possible non-minimal couplings~\cite{Deser:2000dz}.

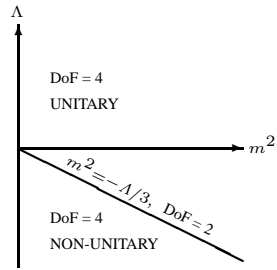
\begin{figure}
$$
\begin{picture}(300,100)(-100,5)
\put(0,100){$\ss \Lambda$}
\put(90,48.5){$\ss m^2$}


\put(20,44){\rotatebox{334}{$\sss m^2=-\L/3 ,\;\;\mbox{\tiny DoF = 2}$}}
\put(15,75){\tiny DoF = 4}
\put(15,65){\tiny UNITARY}
\put(15,22){\tiny DoF = 4}
\put(15,12){\tiny NON-UNITARY}



\put(3,50){\vector(1,0){85}}
\put(3,5){\vector(0,1){92}}

\thicklines

\put(3,50){\line(2,-1){85}}




\end{picture}
$$
\caption{Phase diagram for neutral, massive $s=3/2$ in cosmological 
backgrounds.\label{jam1}}
\end{figure}

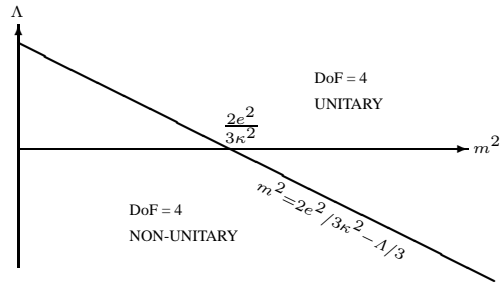
\begin{figure}
$$
\begin{picture}(600,100)(-100,5)
\put(0,100){$\ss \Lambda$}
\put(175,48.5){$\ss m^2$}


\put(92,35){\rotatebox{334}{$\sss m^2=2e^2/3\kappa^2-\L/3$}}
\put(115,75){\tiny DoF = 4}
\put(115,65){\tiny UNITARY}
\put(45,25){\tiny DoF = 4}
\put(45,15){\tiny NON-UNITARY}



\put(80,57){$\sss\frac{2e^2}{3\kappa^2}$}

\put(3,50){\vector(1,0){170}}
\put(3,5){\vector(0,1){92}}

\thicklines

\put(3,90){\line(2,-1){180}}




\end{picture}
$$
\caption{Phase diagram for charged, massive $s=3/2$ coupled
to cosmological Einstein--Maxwell backgrounds.\label{jam2}}
\end{figure}

Our first new result is that the
onset of the unitarity/causality difficulty for massive $s=3/2$ in
pure E/M backgrounds can be traced to a novel gauge invariance of
the timelike component of the Rarita--Schwinger equation at E/M field
strengths tuned to the mass. Although the full system is not
invariant, a consequence of this invariance is signal
propagation with lightlike characteristics. Beyond this tuned
point, {\it i.e.}, for large enough magnetic field $\vec
B^2>(\frac{3m^{2}}{2e})^{\!^{^2}}$ 
(or better, small/large enough mass/charge),
the system is neither causal nor unitary. This is an old result
but its rederivation in terms of a gauge invariance is
edifying.
[Higher spin models in constant curvature backgrounds also
enjoy unexpected gauge invariances at values of the mass tuned
to the background, {\it i.e.} to the cosmological constant. The
flat space limit $m^2>\!\!>\Lambda$ is, of course, always causal,
unitary and massive.] Our second $s=3/2$ result is that in {\it dynamical}
Maxwell--Einstein backgrounds, causality can be maintained for any
choice of E/M field, for certain values of the mass: It is useful, in this
context
to first 
consider the $(m^2,\Lambda)$ phase diagram in Figure~\ref{jam1} for neutral
(Majorana) $s=3/2$ in a cosmological background~\cite{Deser:2001us,
Deser:2001pe}.
The strictly massless line $m^2=-\Lambda/3$ divides the plane
into unitarily allowed and forbidden regions, with the indicated
DoF and corresponds to linearized ${\cal N}=1$ cosmological 
SUGRA~\cite{Townsend:1977qa}. The
addition of an on-shell Maxwell background shifts the line and we
will obtain the new phase diagram depicted in Figure~\ref{jam2}
valid for charged (Dirac)
systems
(the indicated DoF are doubled in going
from Majorana to Dirac when counting real components).  
Hence, causality and unitarity can be
maintained for a charged massive $s=3/2$ field, but only at the
cost of Planck scale masses.

The requirement of Planck scale masses for causal propagation
is unaffected by further non-minimal couplings as
demonstrated by analyzing the characteristic surfaces when a
SUGRA-inspired non-minimal magnetic moment coupling is
added. However, the structure of the $(m^2,\Lambda)$ phase diagram 
(see Figure~\ref{jam3}) 
is richer in this case:
\begin{figure}
$$
\begin{picture}(600,100)(-100,5)
\put(0,100){$\ss \Lambda$}
\put(175,48.5){$\ss m^2$}


\put(20,44){\rotatebox{334}{$\sss m^2=-\L/3$}}
\put(20,75){\tiny DoF = 4}
\put(20,65){\tiny NON-UNITARY}
\put(20,15){\tiny DoF = 4}
\put(20,5){\tiny UNITARY}
\put(110,30){\tiny DoF = 4}
\put(110,20){\tiny UNITARY}
\put(90,-13){\tiny DoF = 4}
\put(90,-23){\tiny NON-UNITARY}




\put(72,56){\rotatebox{90}{$\sss m^2=2e^2/\kappa^2$}}

\put(3,50){\vector(1,0){170}}
\put(3,-25){\vector(0,1){125}}

\put(77.5,8.5){$\bullet\;$}
\put(84,12){\tiny SUGRA}

\thicklines

\put(3,50){\line(2,-1){150}}
\put(80,-25){\line(0,1){125}}




\end{picture}
$$
\vspace{.1cm}
\caption{Phase diagram for charged, massive $s=3/2$ 
with SUGRA-inspired non-minimal coupling to cosmological Maxwell--Einstein
backgrounds.\label{jam3}}
\end{figure}
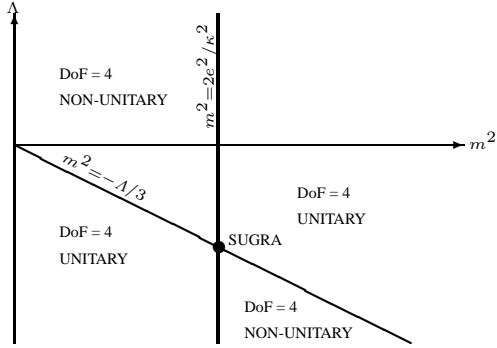
The point $m^2=-3\Lambda=2e^2/\kappa$ corresponds to the
tunings of charge, mass and cosmological constant required for
local supersymmetry of charged cosmological
gravitini
(linearized ${\cal N} =2$ SUGRA~\cite{Freedman:1977aw}). 
Causality is maintained along
the line $m^2=2e^2/\kappa^2$ for any value of $\Lambda$~\cite{Deser:1977uq}. 
For the region around flat space
$\Lambda\approx0$, causality can only be achieved for Planck mass
gravitini, $m^2\geq 2e^2/\kappa^2$, as mentioned above.
Ironically, extremely light, charged gravitini with
$m^2<-\Lambda/3$ and $m^2\leq2e^2/\kappa^2$ do propagate causally. Observe that
the unitarily forbidden region of the neutral $s=3/2$ 
theory~\cite{Deser:2001us,
Deser:2001pe} is split in two in the presence of SUGRA-inspired 
E/M interactions. The new unitary region is not useful for effective
theories of $s=3/2$ excitations since the allowed masses are very small.

It is natural to query whether the rich structure exhibited by 
the 
massive $s=3/2$ system
in Maxwell--Einstein backgrounds
is  generic to higher spins or specifically inherited from ${\cal
N}=2$ cosmological SUGRA. The question is interesting from
a phenomenological standpoint also, since one would like to
develop effective field theoretical methods for relativistic
higher spins interacting with E/M backgrounds.
Therefore, we also perform an analysis of charged $s=2$
and confirm the suspicion that
$s=3/2$ is exceptional and can be viewed as a softly broken version of
${\cal N}=2$ cosmological SUGRA, rather than a predictor of
generic higher spin properties.

Higher spins $\geq2$ actually suffer a much more serious difficulty than
acausality {\it \`a la} $s=3/2$. Manifestly relativistic
descriptions of higher spins demand an ever increasing set of
field components as compared to actual physical DoF. 
Free actions yielding a consistent set of constraints 
eliminating the unphysical components are known for all spins,
massive and massless~\cite{Fronsdal:1978rb, Fang:1978wz,
Curtright:1979uz, deWit:1980pe, Singh:1974qz}. However, the
consistency of these constraints is not guaranteed in the presence
of covariant derivatives associated with gauge interactions. This
problem, observed first in the context of massive charged 
$s=2$~\cite{Federbush:1961}, is that constraints may disappear and the
number of propagating
DoF is  discontinuous  from that of the underlying free theory --
obviously a pathological situation. Although for charged
$s=2$ the DoF problem was cured~\cite{Federbush:1961} in flat
space by introducing a (unique) $g=1/2$
magnetic moment coupling, there is no DoF-preserving gravitational
interaction in general gravitational 
backgrounds~\cite{Aragone:1980bm,Buchbinder:1999ar,Buchbinder:2000fy}. 
The problem is
only compounded as the spin increases, requiring consistency of an
ever larger set of constraints. It is not even clear whether
any DoF preserving coupling to electromagnetic or (not purely
cosmological) gravitational backgrounds exists at all for
$s\geq5/2$.
For $s=2$, this means that there is no analog of the improved
$s=3/2$ causality bounds which relied on dynamically coupled
Maxwell--Einstein backgrounds to use equivalence between Einstein
and E/M stress tensors. Furthermore, the pure E/M $g=1/2$ model 
is acausal~\cite{Kobayashi:1978xd,
Kobayashi:1979mv}. For completeness
we review the computation of its characteristics.

Finally, we consider the separate problem of tree unitarity
for spins $s\geq2$.
Previous authors have suggested that a novel $g=2$ magnetic moment
coupling yields improved tree unitarity properties~\cite{Ferrara:1992yc}. 
In fact, not only does
this coupling fail to preserve the correct DoF count, but as 
we will explicitly show, 
it {\it cannot} yield tree unitary Compton scattering
amplitudes. Taken as a whole, our results highlight the 
difficulties associated
with an effective description of higher spins by local quantum
field theories.

The material is laid out as follows. Section~\ref{3/2}
contains our new causality bounds for the charged gravitating massive
$s=3/2$ system. Minimal, SUGRA-inspired non-minimal and
generic couplings are treated in
Sections~\ref{3/2min}, \ref{3/2sugra} and \ref{3/2nonmin},~respectively.
Spin~2 is dealt with in Section~\ref{spin2}, beginning with a
review of the DoF results of~\cite{Federbush:1961} in
Section~\ref{2model}. Our streamlined acausality proof for E/M
backgrounds may be found in Section~\ref{2causality} and a
detailed discussion of tree unitarity in terms
of the $g$-factor 
is in Section~\ref{2unitarity}.
Our conclusions are summarized in Section~\ref{pig}.

\section{Charged Gravitating Spin $\bf 3/2$}

\label{3/2}

The massive Rarita--Schwinger field $\psi_\m$ is the original
instance of acausality in external E/M backgrounds~\cite{Johnson:1961vt, 
Velo:1969bt}.
It is known to maintain the correct
($2s+1=4$) DoF count so long as its non-minimal coupling is of the
form~\cite{Capri:1972,Deser:2000dz}
 \be {\cal L}_{\rm NM}\sim\psib_\m{\cal F}^{\m\n}\psi_\n\, ,\label{F00} \ee
 where
$\xi_\mu{\cal F}^{\m\n}\xi_\n=0$ for any timelike vector
$\xi_\mu$. It follows that the timelike component $\xi .\psi$ of
$\psi_\mu$ is a Lagrange multiplier (imposing a constraint); 
otherwise terms quadratic in
$\xi.\psi$ appear and imply unwanted,
propagating DoF. In particular,  minimal coupling to gravity and
electromagnetism respects the requirement~\eqn{F00}.

In this Section we study causal propagation for $s=3/2$ in both
E/M and gravitational backgrounds. We take gravity and
electromagnetism to be dynamical, so that the Maxwell stress
tensor is equivalent to the Einstein tensor. [We mostly ignore
back-reaction on the $s=3/2$ probe field itself, but will also
briefly discuss the case where all fields are dynamical.] The main
result is that once gravity is dynamical, $s=3/2$ causality can be
regained at the cost of a Planck scale mass~$m$. The
responsible mechanism is related to the consistency of the underlying $\N=2$
(broken) SUGRA theory. We also show that the original
E/M acausality is caused by a novel
gauge invariance of certain field equations.

\subsection{Minimal Coupling}

\label{3/2min}

The massive $s=3/2$ Lagrangian minimally coupled to both gravity and
electromagnetism is\footnote{Our metric is ``mostly plus'' and Dirac
matrices are in turn ``mostly hermitean'', in particular
$\g^5=\g^5{}^\dagger=-i\g^{0123}=\frac{i}{24\sqrt{-g}}\,\g_{\m\n\r\s}
\e^{\m\n\r\s}$. We (anti)symmetrize with unit weight.}
\begin{equation}
{\cal L}
=-\sqrt{-g}\; \psib_\m\,\g^{\m\n\r}\,\D_\n\psi_\r\, ,
\label{wakno}
\end{equation}
where the mass term is included in the extended covariant
derivative,
\be \D_\m\equiv D_\m+\frac{m}{2}\,\g_\m \, ,\ee 
satisfying $[\D_\m,\D_\n]=[D_\m,D_\n]+\frac{m^2}{2}\,\g_{\m\n}$. 
The usual covariant derivative on the Rarita--Schwinger field reads 
\be
D_\m\psi_\n=
 \d_\m\psi_\n-\G^\r{}_{\m\n}\,\psi_\r
+\frac{1}{4}\,\w_{\m mn}\g^{mn}\psi_\n
+ieA_\m\psi_\n\, , 
\ee
with commutator 
\be
[D_\m,D_\n]\,\psi_\r=
-R_{\m\n\r}{}^\s(g)\,\psi_\s+\frac{1}{4}\,R_{\m\n mn}(\w)\g^{mn}\,\psi_\r
+ie\,F_{\m\n}\psi_\r\, .
\ee
We generally drop the labels $g$ and $\w$ with the curvature
convention 
$R_{\m\n\r\s}\equiv R_{\m\n\r\s}(g)=-e_\r{}^ae_\s{}^bR_{\m\n ab}(\w)$. 
The Rarita--Schwinger equation (following from \eqn{wakno})~is
\be
R^\m\equiv \g^{\m\n\r}\D_\n\psi_\r=0\, .
\label{RS}
\ee

We first note that causality is equivalent to the consistency of
constraints for all values of the background fields. Along the
lines of~\cite{Deser:2001us,Deser:2001pe}, 
we can study the problem by searching for gauge
invariances: Consider the case of a pure E/M background and search
for a gauge invariance
\be 
\delta\psi_\m=\D_\m \varepsilon\, .\label{latte}
\ee
Varying~\eqn{RS}, we find, in terms of the dual $\wt{F}^{\m\n}
\equiv \frac{1}{2} \, \epsilon^{\mu\nu\alpha\beta} F_{\alpha\beta}$,
\be 
\delta R^\m=-\Big(e\g^5 \wt F^{\m\n}\g_\n+
\frac{3m^2}{2}\g^\m\Big)\,\varepsilon\equiv\Pi^\m\,\varepsilon\, .\label{rocky}
\ee
It is easy to verify that there is no simultaneous eigenspinor $\varepsilon$ 
of the operator $\Pi^\mu$ with vanishing eigenvalue for each value of the 
Lorentz index $\mu$: the massive Rarita--Schwinger equation is not gauge
invariant under~\eqn{latte}. However, if we just take the timelike
component $\mu=0$ and examine the determinant of the matrix multiplying
$\varepsilon$ in the variation~\eqn{rocky} we find
\be 
\det \Pi^0=e^4\,\Big(\vec B^2-\Big[\frac{3m^2}{2e}\Big]^2\Big)^2\, .
\ee
Hence when
\be 
B\equiv|\vec B|=\frac{3m^2}{2e} \label{machiatto}
\ee
and $\varepsilon$ is an eigenspinor of $\Pi^0$ with vanishing
eigenvalue, the transformation~\eqn{latte} is an invariance of the
timelike component of the Rarita--Schwinger equation.
Although~\eqn{latte} is not an invariance of the full field
equations (or action), 
this suffices to ruin the consistency of the constraints
and permit luminal signal propagation. The bound~\eqn{machiatto}
on the magnetic field is precisely the one discovered in~\cite{Johnson:1961vt, 
Velo:1969bt}
and occurs when the mass is tuned to the
external background, just as happened when $m^2$ was
tuned to the constant curvature gravitational 
background~\cite{Deser:2001us,Deser:2001pe}.

Next we perform an explicit analysis of the system's
characteristics using the method first introduced in this context
in~\cite{Madore:1973,Madore:1975}. By studying a shock whose first
derivative is discontinuous across the wavefront, we may determine
the maximum speed of propagation. The leading discontinuities across
the characteristic (or wavefront) $\Sigma$ are denoted by square
brackets \be [\d_\m\psi_\n]_{_{\scriptstyle \Sigma}}
\equiv\xi_\m\Psi_\n \, ,\label{romeo} \ee where $\xi_\mu$ is a vector
normal to the characteristic and $\Psi_\nu$ is some non-vanishing
vector-spinor defined on the characteristic surface. Propagation
is acausal whenever the field equations admit characteristics with
timelike $\xi_\mu$.
Examining the discontinuity in the field equation~\eqn{RS} and its
gamma-trace we find
\bea
{}[R_\m-\frac{1}{2}\,\g_\m\g.R]_{_{\scriptstyle \Sigma}} 
&=&\Sl{\xi}\,\Psi_\m-\xi_\m\g.\Psi\;=0\, ,\\
{}[\g.R]_{_{\scriptstyle \Sigma}}
&=&2\,(\xi\!\!\!/\,\g.\Psi-\xi.\Psi) \;=\;0\, , \eea which combine to
give \be \xi^2\,\Psi_\m=\xi_\m\,\xi.\Psi\, . \label{juliet} \ee
Clearly there are two possibilities: The first is vanishing 
$\xi.\Psi$ and $\xi^2$; the maximal speed of propagation is then
governed by the light-cone and the model is causal. Alternatively
$\xi.\Psi\neq0$; we proceed by contradiction and assume that
$\xi_\mu$ is timelike $(\xi^2=-1)$:
To determine when a timelike normal  vector to the characteristic
is allowed, we consider the secondary (Lagrangian) constraint
\be \D.R=-\frac{3}{2}\,
m^2\g.\psi +e\,\g^5\g.\wt F.\psi -\frac{1}{2}\,\g.G.\psi\, .
\label{turnips} 
\ee
 Taking a further derivative and computing the
discontinuity we learn
\be \xi^\m\,[\d_\m \D.R]_{_{\scriptstyle
\Sigma}} =-\frac{3}{2}\, m^2\;
\g.\Big[\,\xi+\frac{1}{3m^2}\,G.\xi+ \frac{2e}{3m^2}\,\g^5\wt
F.\xi\,\Big]\;\xi.\Psi \, . 
\ee
The model is acausal
if~\eqn{turnips} admits a solution for non-zero $\xi.\Psi$. Since
for any pair of vectors $(v_\m,a_\m)$, $\det[\g^\m (v_\m+\g^5
a_\m)]$ $\equiv$ $(v+a)^2\,(v-a)^2$, we are led to study the
solubility of
\bea \lefteqn{ \xi^2\,+\,\frac{2}{3m^2}\,\xi.G.\xi
\,+\,\frac{1}{(3m^2)^2}\,\xi.G.G.\xi\qquad}&&\nn\\&&
\pm\;\Big(\frac{2}{3m^2}\Big)^2\,e\,\xi.G.\wt F.\xi
\,-\,\Big(\frac{2e}{3m^2}\Big)^2\,\xi.\wt F.\wt F.\xi\;=\;0 \; .
\label{drive} \eea
 Without loss of generality we take
$\xi_\m=(1,0,0,0)$ and impose Einstein's equations on the
background fields
\be
G^{00}=-\k^2 T^{00}=-(\k^2/2)(\vec E^2+\vec B^2)\,,\qquad
G^{0i}=-\k^2 T^{0i}=-\k^2(\vec E\times \vec B)^i\, .
\ee
The condition~\eqn{drive} then becomes (the $\pm$ term
vanishes on-shell)
\be
(1+\E+\B)^2\,-\,4\,\E\B\,\sin^2\theta-2Q\,\B=0\, ,
\label{ben}
\ee
$$
Q\equiv \frac{4e^2}{3\k^2 m^2}\,,\qquad 
\E \equiv\frac{\k^2\vec E^2}{6m^2} \; , \qquad
\B\equiv\frac{\k^2\vec B^2}{6m^2}
$$
($\theta$ is the angle between
$\vec E$ and $\vec B$). When the gravitational interaction is
turned off ($\k=0$) only the terms $1-2Q\,\B$ survive and yield
the original causality bound~\cite{Velo:1969bt} on the
magnetic field, $\vec B^2<(\frac{3m^2}{2e})^{\!^2}$. However for any
$\k\neq0$, there
are values of ($e$,$m$) such that~\eqn{ben} has no solutions
for any E/M field strength $F_{\m\n}$, and the model is causal.
This is the case when the bound\footnote{ Consider the l.h.s.
of~\eqn{ben} as a quadratic in $\B$ (vanishing $\B$ is always
causal) and require that the discriminant be negative, which
yields $Q<2+2\E\,\cos^2\theta$, whose most stringent bound on $Q$
in~\eqn{jerry} is attained for $\vec E \perp \vec B$.} 
\be Q<
2\;\Longleftrightarrow \; m^2 > \frac{2}{3}\;\frac{\,e^2\,}{\k^2}
\; \label{jerry} 
\ee 
is satisfied. Since masses of this Planckian
order would invalidate our no back-reaction assumption, we prefer
to state our result as follows: massive $s=3/2$ minimally coupled
to gravity and electromagnetism cannot propagate causally for mass
to charge ratios smaller than the Planck mass. We will discuss the
extension to back-reaction at the end of this Section.

The addition of a cosmological constant term shifts the mass term 
$m^2\!$~$\!\rightarrow\!$~$m^2\!$~$\!+\!$~$\!\L/3$ and modifies the bound to
read $m^2Q<2(m^2+\Lambda/3)$ so that
\be
m^2>\frac{2e^2}{3\k^2}-\frac{\L}{3}\, .
\ee
These results yield the $(m^2,\Lambda)$ phase diagram of Figure~\ref{jam2}.


\subsection{SUGRA Causality}

\label{3/2sugra}

$\N=2$ AdS SUGRA~\cite{Freedman:1977aw} describes a
Dirac gravitino coupled (both minimally and non-minimally) to
electromagnetism and gravity with cosmological constant. 
The apparent mass term for the gravitino has precisely the coefficient
required to ensure true
masslessness in the presence of a tuned cosmological 
constant~\cite{Deser:1977uq}. However,
supersymmetry can 
be softly broken  by turning off the cosmological constant
{\it without} losing
causal propagation~\cite{Deser:1977uq}, {\it i.e.}, keeping the 
$s=3/2$ mass term and detuning the cosmological constant
from its supersymmetric value. This implies 4 massive $s=3/2$ DoF
and,
in effect, yields a consistent (albeit very heavy) charged massive
$s=3/2$ model.

One might wonder whether this mechanism can be employed to derive
consistent phenomenologically applicable models. Unfortunately the
result of~\cite{Deser:1977uq} also requires a Planck mass
gravitino: Although the cosmological constant term was removed,
the gravitino mass parameter kept its supersymmetrically tuned
value $m=\sqrt{2}\,eM_P\equiv\sqrt2\,e/\kappa$.
Here we
study the more general case where both the gravitino mass and
cosmological constant terms are detuned. The final result is,
however, similar to that of~\cite{Deser:1977uq}: Causality
requires the $s=3/2$ mass to be either so large or so small
that the model cannot be applied to physical situations. In the causal large 
mass regime, the lower bound on the mass is the supersymmetric value
$m>\sqrt2\,e/\kappa$, whilst causal small mass models require
an AdS background with an upper mass bound $m<\sqrt{-\L/3}$.

In this model, we  include the non-minimal E/M coupling required by
$\N=2$ AdS SUGRA,
\be {\cal L}_{\rm
Non-Min}=-\frac{ie}{m}\,\sqrt{-g}\,\psib_\m\F^{\m\n}\psi_\n\, ,
\label{aggy} 
\ee
where
\be
\F^{\m\n}=-F^{\m\n}_+\equiv-(F^{\m\n}+i\g^5 \wt F^{\m\n})\, .
\ee
Again we assume that the Rarita--Schwinger field is a probe moving in
a combined gravitational and E/M background which satisfies the bosonic field
equations of $\N=2$ AdS SUGRA
\be
D_\m F^{\m\n} = 0 = G^{\m\n}+\k^2\,T^{\m\n}-\Lambda g^{\m\n}\, .
\label{terence}
\ee

Once again, we proceed by contradiction, {\it i.e.} $\xi^2=-1$, since
the leading discontinuities still satisfy~\eqn{romeo}--\eqn{juliet},
but the crucial secondary constraint~\eqn{turnips} is now modified as the 
Rarita--Schwinger field equation reads
\be
R^\m=\g^{\m\n\r}\D_\n\psi_\r-\frac{ie}{m}F^{\m\n}_+\psi_\n=0\, .
\ee
To compute the secondary constraint $\D.R$ we first
note that
\be
\D.F_+.\psi=F_+^{\m\n}\D_\m\psi_\n-im\g^5\g.\wt F.\psi\, .
\ee
Furthermore
\be
\g^{[\m}R^{\n]}=2\,\D^{[\m}\psi^{\n]_{\!_{_+}}}
-\g^{[\m} \Big((\D\!\!\!\!/-m) \psi^{\n]} - (\D^{\n]}-m\g^{\n]})\g.\psi\Big)
-\frac{ie}{m}\,\g^{[\m}F^{\n]\rho}_+\psi_\r=0\, ,
\ee
where we have introduced the notation 
$X^{[\m\n]_{\!_{_+}}}\equiv X^{[\m\n]}+i\g^5 \wt
X^{\m\n}$ 
(note that 
$X^{[\m\n]_{\!_{_+}}}Y_{\m\n}=
(1/2)\,X^{[\m\n]_{\!_{_+}}}Y_{[\m\n]_{\!_{_+}}}$).
The terms in brackets are equal to (see equation (32) of~\cite{Deser:2000dz})
$(ie/2m)$ $\g_\r\g^\m F_+^{\r\s}\psi_\s=(ie/m)\, F_+^{\m\r}\psi_\r$
using the identity $\g.F_+.\psi=0$. Hence 
\be
\D^{[\m}\psi^{\n]_{\!_{_+}}}=\frac{ie}{m}\,\g^{[\m}F_+^{\n]\r}\psi_\r
\ee
and in turn 
\be
\D.F_+.\psi=-im\g^5 \g.\wt F.\psi -\frac{ie}{m}\,\g.T.\psi\, .
\label{hash}
\ee
Here we used $(F_-.F_+)_{\m\n}=(F.F+\wt F .\wt F)_{\m\n}=-2\,T_{\m\n}$
(with $F_-^{\m\n}\equiv F^{\m\n}-i\g^5\wt F^{\m\n}$). 
The first term on the right hand side of~\eqn{hash} precisely cancels
the troublesome second term in~\eqn{turnips} and we find
\be
\D.R=-\frac{3}{2}\, m^2 \g.(\psi +\frac{1}{3m^2}\, G.\psi +
\frac{2e^2}{3m^4}\, T.\psi)\, .
\ee
For the discontinuity in the derivative of the secondary constraint we
find (employing the background field equations~\eqn{terence})
\be
\xi^\m[\d_\m \D.R]_{_{\scriptstyle \Sigma}} =
-\frac{3}{2}\, m^2 \g.\Big[\xi\,(1+\frac{\Lambda}{3m^2})
+T.\xi\,(\frac{2e^2}{3m^4}-\frac{\k^2}{3m^2})\,\Big]\,\xi.\Psi\, .
\ee
Firstly note how the result of~\cite{Deser:1977uq} is recovered:
The SUGRA tuning of the $s=3/2$ mass
\be
m^2=\frac{2e^2}{\k^2}
\label{suggy}
\ee
maintains causality even when $\L=0$ (and in fact for any
cosmological constant). The point where in addition,
$\Lambda=-3m^2$ corresponds to unbroken 
$\N=2$ SUGRA whose $s=3/2$ field is genuinely massless with null propagation. 

To study causality more generally,
we require the determinant of the matrix multiplying $\xi.\Psi$
 not to vanish. We now assume that the equality~\eqn{suggy} does
not hold. Again, by contradiction put $\xi_\m=(1,0,0,0)$. Calling 
$\alpha \equiv 1+\Lambda/3m^2$ and 
$\E\equiv (\k^2/2)(2e^2/3m^4-\k^2/3m^2)\vec E^2$ and
$\B\equiv (\k^2/2)(2e^2/3m^4-\k^2/3m^2)\vec B^2$ 
this determinant 
vanishes iff
\be
(\alpha-\E)^2-2(\alpha-\E+2\sin^2\!\theta\,\E)\,\B+\B^2=0\, ,
\ee
which implies the model is causal whenever the discriminant is
negative, namely when
\be
\sin^2\!\theta\,\E(\alpha-\cos^2\!\theta\,\E)<0\, .
\ee
There are two cases. Firstly if 
\be
m^2>\frac{2e^2}{\k^2}\, ,
\ee
then $\E<0$ and the model {\it is}  causal so long as $\alpha>0$ which
implies the additional restriction on the mass
\be
m^2>-\Lambda/3\, .
\ee
Alternatively if
\be
m^2<\frac{2e^2}{\k^2}\, ,
\ee
then $\E>0$ and the model is causal only if $\alpha<0$,
{\it i.e.} for masses
\be
m^2<-\L/3\, .
\ee
The corresponding $(m^2,\L)$ phase diagram is that of Figure~\ref{jam3}.

\subsection{Non-minimal Couplings and Dynamical
Rarita--Schwinger Fields}

\label{3/2nonmin}

We end this Section by considering the effect of general non-minimal
couplings on the above results along with some comments on  
dynamical fermi fields.

In principle it is possible to add the most general non-minimal
couplings along the lines of~\cite{Deser:2000dz}; however as shown
there, the most general magnetic moment couplings are already mapped
out by non-minimal interactions of the form
\be
{\cal L}_{\rm Non-Min}=-\frac{iel}{m}\,\sqrt{-g}\;\psib.\F.\psi\, .
\label{laggy}
\ee
Indeed, the parameter $l$ may be expressed in terms of the gyromagnetic
ratio as~\cite{Deser:2000dz}
\be
l=\frac{3g-2}{4}\, .
\ee 
The supersymmetric case $l=1$ corresponds to $g=2$.

The discontinuity in the derivative of the
secondary constraint is modified to read
\bea
\lefteqn{
\xi^\m[\d_\m \D.R]_{_{\scriptstyle \Sigma}} \;=\;}&&\nn\\&&
-\frac{3}{2}\, m^2 \g.\Big[\Big(1-\frac{\Lambda}{3m^2}\Big)\,\xi
+\frac{2e(1-l)}{3m^2}\,\g^5\wt F.\xi
+\Big(\frac{2e^2l^2}{3m^4}-\frac{\k^2}{3m^2}\Big)\,T.\xi\,\Big]\,\xi.\Psi\, .
\nn\\
\eea
The analysis now proceeds in a fashion similar to that presented
above. For simplicity we concentrate on the most physically
relevant case, namely $\L=0$ and $m<\sqrt2e/\k$. It easily verified 
that the relevant discriminant is a sum of
squares (saturated at the
supersymmetric point $l=1$, $m=\sqrt{2}e/\k$) 
\be
\Big(\frac{2e^2l^2}{m^2}-\k^2\Big)^2+\Big(2e(1-l)\Big)^2\geq0
\ee
and can never be negative: non-minimal couplings cannot restore
causality in the flat space, small mass regime.

Finally we consider the consequences of making the Rarita--Schwinger 
field dynamical.
This requires the inclusion of the
$s=3/2$ stress-energy tensor on the right hand side of Einstein's
equations. In the case of SUGRA, the additional trilinear terms
in the field equations from varying the four fermi terms required for
local supersymmetry of the action will cancel the new contributions
from the $s=3/2$ stress-energy tensor. For minimal coupling,
however, one simply has to deal with these new terms. For example, if
one views the classical value of spinor bilinears as an expectation
value and assumes that $T_{3/2}^{00}>0$ then all
additional
contributions to the polynomial~\eqn{ben} are at least positive and
will not weaken the resulting causality bounds.







\section{Spin~2}

\label{spin2}

We now analyze the first instance in which there is no underlying
protection, in contrast to that provided by SUGRA for $s=3/2$: while 
$s=2$ is a tensor theory, when charged it bears little
resemblance to its one conceivable relative, Einstein gravity. 
Charged massive $s=2$ preserves the correct DoF
only for gyromagnetic ratio $g=1/2$, but  even this theory
suffers from the usual causality difficulties. Furthermore, it has
no good DoF coupling to general gravitational backgrounds, so there
is no analog of the causality bounds found for $s=3/2$.

\subsection{DoF Count}

\label{2model}

The unique charged massive $s=2$ theory maintaining the
correct $2s+1=5$ (complex) DoF count in E/M backgrounds was 
discovered in~\cite{Federbush:1961}. To establish notation, we first write
the free Lagrangian,
\be
{\cal L}=\frac{1}{2}\,\phi^*{}^{\m\n}\,G_{\m\n}(\d,\phi)
-\frac{1}{2}\,m^2\,\Big(\phi^*{}^{\m\n}\phi_{\m\n}-\phi^*\phi\Big)\, ,
\ee
where $\phi\equiv\phi_\m{}^\m$. The linearized Einstein tensor 
\be
G_{\m\n}(\d,\phi)=\Box \,(\phi_{\m\n}-\eta_{\m\n}\phi)+\d_\m\d_\n\phi
+\eta_{\m\n}\d.\d.\phi-2\d.\d_{(\m}\phi_{\n)}
\, ,\quad\d.\phi_\m\equiv\d^\r\phi_{\r\m}
\label{Einstein}
\ee
satisfies the
Bianchi identity
\be\d.G_\m=0\, .\ee
The free field equations are
\be
{\cal G}_{\m\n}\equiv
(\Box-m^2) (\phi_{\m\n}-\eta_{\m\n}\phi)+\d_\m\d_\n\phi
+\eta_{\m\n}\d.\d.\phi-2\d.\d_{(\m}\phi_{\n)}=0
\label{laddy}
\ee
from which the usual on-shell conditions for  massive $s=2$, 
\be(\Box-m^2)\phi_{\m\n}=0=\d.\phi_\m=\phi\, ,\ee
follow upon using 
\be
\d.\d.{\cal G}+(m^2/2)\,{\cal
G}_\m{}^\m=(3m^4/2)\,\phi=0=\d.{\cal G}_\m=-m^2(\d.\phi_\m-\d_\m\phi)\, .
\ee

The minimal coupling procedure $\d_\m\rightarrow D_\m=\d_\m+ie
A_\m$ is ambiguous since possible reorderings of partial derivatives
in the last term of~\eqn{laddy} lead to a
one parameter  family of couplings (presciently labelled by the gyromagnetic 
ratio $g$) 
\be
\partial.\partial_{(\m}\phi_{\n)}\longrightarrow
g\, D.D_{(\m}\phi_{\n)}+(1-g)\,D_{(\m}D.\phi_{\n)}= D_{(\m}D.\phi_{\n)} 
+ieg F_{\r(\m}\phi_{\n)}{}^\r\, .
\label{nrat}
\ee
[The particular choice $g=1$ corresponds to minimal coupling in a 
first order formalism in which\footnote{See~\cite{deWit:1980pe,Damour:1987vm}
for a geometric formulation of free massless higher spin fields
in terms of generalized Christoffel symbols.}
\be
-\frac{1}{2}\,
G_{\m\n}=\d_\r \Omega^\r{}_{\m\n}-\d_{(\m}\Omega^\r{}_{\n)\r}
-\frac{1}{2}\,\eta_{\m\n}(\d_\r\Omega^{\r\s}{}_\s-\d_\r\Omega^{\s\r}{}_\s)
\ee
and the on-shell linearized Christoffel symbols are
\be
\Omega_{\r\m\n}=\d_{(\m}\phi_{\n)\r}-\frac{1}{2}\,\d_\r\phi_{\m\n}\;.\Big] 
\ee 

To study DoF in an E/M background, we begin with the one parameter family
of minimally coupled Lagrangians
\bea
{\cal L}&=&-\frac{1}{2}\,D^\r\phi^*{}^{\m\n}D_\r\phi_{\m\n}
-\frac{1}{2}\,m^2\phi^*{}^{\m\n}\phi_{\m\n}
+\frac{1}{2}\,D^\r\phi^* D_\r\phi+\frac{1}{2}\,m^2\phi^*\phi
\nn\\&&
-\frac{1}{2}\,D.\phi^*{}^\m D_\m\phi
-\frac{1}{2}\,D^\m\phi^*D.\phi_\m+D.\phi^*{}^\m D.\phi_\m
+ie\,g\,\phi^*{}^{\r\m}F_{\m\n}\phi^\n{}_\r\, .\nn\\
\eea
As always with models involving constraints, it is necessary to
first verify that interactions maintain the correct DoF.
Here we must investigate the single and double divergences 
of the field equation
\bea
{\cal G}_{\m\n}&=&(D^2-m^2) (\phi_{\m\n}-\eta_{\m\n}\phi)+D_{(\m}D_{\n)}\phi
+\eta_{\m\n}D.D.\phi-2D_{(\m}D.\phi_{\n)}
\nn\\&&
-2ie\,g\,F_{\r(\m}\phi_{\n)}{}^\r=0\, .
\eea
We now examine candidate constraints arising 
from single and double divergences of the field equations,
recast in terms of Lichnerowicz wave operators in the Appendix as
\bea
{\cal G}_{\m\n}&=&(\DL{2}-m^2) (\phi_{\m\n}-\eta_{\m\n}\phi)+D_{(\m}D_{\n)}\phi
+\eta_{\m\n}D.D.\phi-2D_{(\m}D.\phi_{\n)}
\nn\\&&
+2ie\,(2-g)\,F_{\r(\m}\phi_{\n)}{}^\r=0\, .
\eea
To begin with, the single divergence
\bea
D.{\cal G}_\m&=&-m^2(D.\phi_\m-D_\m\phi)
\nn\\
&&+\;ie\,\Big\{\,
\frac{1}{2}\,j_\m\phi-(1-g)\, j.\phi_{\m}+g\,(\d_\r F_{\m\n})\phi^{\n\r}
\nn\\&&
+\,F_{\m\n}\,\Big[(1+g)\,D.\phi^\n-\,\frac{3}{2}\,D^\n\phi\Big]\,
+(2-g) F^{\n\r}D_\r\phi_{\m\n}
\,\Big\}
\label{smod}
\eea
involves only first derivatives even in an E/M background with arbitrary
choice of non-minimal couplings and is
therefore a (Lagrangian) constraint. However, the double divergence
\bea
D.D.{\cal G}+\frac{1}{2}\,m^2\ {\cal G}_\m{}^\m&=&\frac{3}{2}\;m^4\,\phi
-ie\,(1-2g)\,F^{\m\n}D_\m D.\phi_\n\nn\\
&+&\!\!ie\,\Big\{\;
2\,g\,j.D.\phi-j.D\,\phi \nn\\
&&\;\;-\,(1-2g)\,(\d_\m j_\n)\phi^{\m\n}
-2\,(1-g)\,(\d_\m F^\r{}_\n) D_\r\phi^{\m\n}\Big\}
\nn\\
&+&\!\!e^2\,\Big\{\;
3\,T_{\m\n}\phi^{\m\n}+(1+g)\,F_{\m\r}F^\r{}_\n\phi^{\m\n}
\Big\}
\label{bouble}
\eea
(where $T_{\m\n}=-F_{\m}{}^\r F_{\r\n}+\frac{1}{4}\eta_{\m\n}F^2$)
is crucial since it includes a term with an E/M field strength multiplying 
double time derivatives of fields
\be
-ie\,(1-2g)\,F^{\m\n}D_\m D.\phi_\n\, .
\ee
This term implies a DoF breakdown 
since a constraint of the free model has become a propagating field equation.
In particular, the leading time derivatives in~\eqn{bouble} are
$-ie$ $(1-2g)$ $F^{0i}\ddot\phi_{0i}$ so that
$F^{0i}\phi_{0i}$ becomes an additional propagating degree of freedom
unless $g=1/2$. 
Therefore only the minimally coupled model with $g=1/2$ maintains the
correct degree of freedom count.  Curiously enough, this is the model 
that takes the average of the ambiguous 
terms\footnote{The coupling
$
{\cal L}_{\rm Non-Min}\propto\frac{ie}{2}\,\phi^*{}^{\r\m}F_{\m\n}\phi^\n{}_\r
$
is the unique non-minimal term not involving derivatives on the matter fields.
It is also not difficult to demonstrate that no coupling of the schematic form
$\phi F D^2 \phi$ involving two derivatives and an E/M field strength
preserves the DoF count.}~\eqn{nrat}.

So far, we have seen that charged massive $s=2$ is uniquely described 
by the (local) Lagrangian
\bea
{\cal L}&=&-\frac{1}{2}\,D^\r\phi^*{}^{\m\n}D_\r\phi_{\m\n}
-\frac{1}{2}\,m^2\phi^*{}^{\m\n}\phi_{\m\n}
+\frac{1}{2}\,D^\r\phi^* D_\r\phi+\frac{1}{2}\,m^2\phi^*\phi
\nn\\&&
-\frac{1}{2}\,D.\phi^*{}^\m D_\m\phi
-\frac{1}{2}\,D^\m\phi^*D.\phi_\m+\frac{1}{2}\,D.\phi^*{}^\m D.\phi_\m
+\frac{1}{2}\,D^\r\phi^*{}^{\m\s}D_\s\phi_{\m\r}\, ,\nn\\
\eea
with field equations
\bea
{\cal G}_{\m\n}&=&(D^2-m^2) (\phi_{\m\n}-\eta_{\m\n}\phi)+D_{(\m}D_{\n)}\phi
+\eta_{\m\n}D.D.\phi\nn\\&&
-D_{(\m}D.\phi_{\n)}-D.D_{(\m}\phi_{\n)}\,=\,0\, .
\eea
Correct DoF are ensured by the five constraint equations
\bea
D.{\cal G}_\m&=&-m^2(D.\phi_\m-D_\m\phi)
\nn\\
&&+\,\frac{ie}{2}\,\Big\{
\,3 F^{\n\r}D_\r\phi_{\m\n}-j.\phi_{\m}
\nn\\&&\qquad
-\,3\,F_{\m\n}D^\n\phi\,+\,j_{\m}\phi
\nn\\&&\qquad
+\,3 F_{\m\n}D.\phi^\n+(\d_\r F_{\m\n})\phi^{\n\r}
\,\Big\}\,=\,0\, ,
\label{vekki}
\\
D.D.{\cal G}+\frac{1}{2}\,m^2\ {\cal G}_\m{}^\m&=&
\frac{3}{2}\,(m^4-\frac{e^2}{2}\,F^2) \,\phi
+\frac{3e^2}{2}\,T_{\m\n}\phi^{\m\n}
\nn\\&&
+\,ie\,(j.D.\phi-j.D\,\phi)
+\,ie\,(\d_\m F_{\n\r})D^\r\phi^{\m\n}\,=0\, .\nn\\
\label{scally}
\eea
Of course a correct DoF count does not yet ensure consistency of the
model since one still has to verify that the above
constraints
eliminate the unphysical DoF for
all points in E/M field space. Equivalently, one can check the 
causality of the model, our next subject.

\subsection{Spin~2 Acausality}

\label{2causality}

Charged massive $s=2$ has long been known to propagate 
acausally~\cite{Kobayashi:1978xd,Kobayashi:1979mv}. For completeness
we provide a streamlined proof using the method of characteristics.
Once again we examine the leading discontinuities of a shockwave,
which are now second order and denoted as
\be
[\d_\m\d_\n\phi_{\r\s}]_{_{\scriptstyle \Sigma}} =\xi_\m\xi_\n\Phi_{\r\s}\, .
\ee
From the field equation and its trace we learn that 
\bea
[{\cal G}_{\m\n}-\frac{1}{2}\,
\eta_{\m\n}{\cal G}_\r{}^\r]_{_{\scriptstyle \Sigma}} &=&
\xi^2\Phi_{\m\n}+\xi_\m\xi_\n\Phi-2\,\xi_{(\m}\xi.\Phi_{\n)}\,=\,0\, ,
\label{maroon}\\
{}[{\cal G}_\m{}^\m]_{_{\scriptstyle \Sigma}} &=&-2\,\xi^2\Phi+2\,\xi.\xi.\Phi\,\,=\,0 \, ,
\label{chatreuse}
\eea
($\Phi\equiv\Phi_\m{}^\m$).
We now study the system for acausal timelike normal vector
$\xi^2=-1$.
Note that since $\Phi_{\m\n}\neq0$, we deduce that
$V_\m\equiv\xi.\Phi_\m\neq0$ (otherwise $\xi^2=0$ and the model would be
causal). So we now impose
\be
\Phi_{\m\n}=-\xi_\m\xi_\n\,\xi.V-2\,\xi_{(\m}V_{\n)}\, ,\qquad \Phi=-\xi.V
\ee
and study further constraints. In particular, the single divergence
con\-st\-raint~\eqn{vekki} gives
\be
\xi^\m[\d_\m D.{\cal G}_\n]_{_{\scriptstyle \Sigma}} =
m^2(V_\n+\xi_\n\,\xi.V)
+\frac{3ie}{2}\,\Big(
F_{\n\r}V^\r
-\xi^\r F_{\r\n}\,\xi.V+\xi_\n\,\xi.F.V
\Big)\, ,
\ee
so that
\be
\Pi_{\n\r}\Big[m^2\eta^{\r\s}+\frac{3ie}{2}\,F^{\r\s}\Big]\,
\Pi_{\s\tau}\,V^\tau=0\, ,
\label{traceless}
\ee
where the projector $\Pi_{\m\n}\equiv\eta_{\m\n}+\xi_\m\xi_\n$.
It is sufficient to search for acausalities in constant background
E/M fields, so with this restriction the
double divergence constraint gives
\be
\xi^\m\xi^\n\,[\d_\m\d_\n(D.D.{\cal G}+\frac{1}{2}\,m^2{\cal
G}_\r{}^\r)]_{_{\scriptstyle \Sigma}} =
-\frac{3}{2}\,(m^4-\frac{e^2}{2}\,F^2+e^2\,\xi.T.\xi)\, \xi.V
-3e^2\,\xi.T.V\, .
\label{trace}
\ee
Causality requires that the system of equations~\eqn{traceless} 
and~\eqn{trace} have no non-zero solution for $V_\m$.
In general~\eqn{traceless} implies the vanishing of the components of
$V_\m$ orthogonal to $\xi_\m$ and~\eqn{trace} in turn removes the
parallel components. However if, regarded as a matrix equation in
the orthogonal subspace, equation~\eqn{traceless} fails to remove the
orthogonal components of $V_\m$ the model will be acausal. The
determinant in this subspace vanishes whenever
\be
\vec B^2=\left(\!\frac{2m^2}{3e}\!\right)^2\, .
\ee
This is the result of~\cite{Kobayashi:1978xd,Kobayashi:1979mv}. 
Requiring that~\eqn{trace} be non-degenerate
yields a different and weaker bound $\vec B^2=(2m^4+\vec E^2)/(3e^2)$.
These differing bounds for the propagation of helicities zero and one
are remiscent of the behaviour found for $s=2$ in cosmological 
backgrounds~\cite{Deser:2001us,Deser:2001pe}.

There is no analog for $s=2$ of the $s=3/2$ 
improved causality behaviour in the presence
of general gravitational backgrounds since the 
model cannot be coupled to gravity whilst maintaining the correct 
DoF~\cite{Buchbinder:1999ar,Buchbinder:2000fy}.
Essentially the most general double divergence constraint always becomes a
propagating field equation even once non-minimal couplings are added.
Clearly, the combination of dynamical Maxwell and Einstein 
fields is unlikely
to yield an
improvement of the situation since the possible double derivative
terms in the charged $s=2$ double divergence constraint are not
proportional to the E/M stress tensor. Furthermore,
in~\cite{Buchbinder:1999ar,Buchbinder:2000fy} consistent 
gravitational backgrounds
have been found but they require that the traceless part of the Ricci tensor
vanish, prohibiting a (trace-free) Maxwell
stress tensor: 
Unlike the $s=3/2$ case, there is no underlying
charged theory such as SUGRA to ensure consistency.

\subsection{Gyromagnetic Ratio and Tree Unitarity}

\label{2unitarity}

A very different requirement that has been imposed on higher spin theories is
tree unitarity~\cite{Ferrara:1992yc}. The high energy behaviour of
partial wave amplitudes is subject to unitarity bounds:
$N$-point amplitudes should grow no faster than $E^{4-N}$ at high
energies $E$. This requirement was used to analyze
the uniqueness of spontaneously broken gauge field theories
as a fundamental description of massive $s=1$ 
particles~\cite{LlewellynSmith:1973ey,Cornwall:1974km}. 
[It is also important, although not sufficient, 
for renormalizability when higher loop diagrams are built from trees.]
The precise mechanism is that terms proportional to
inverse masses in  propagators for spins $\geq1$ generate a gauge
transformation at the vertices. These terms are therefore 
cancelled by requiring vertices to
satisfy gauge Ward identities.
However, although this mechanism can be applied successfully to 
spins~$<2$~\cite{LlewellynSmith:1973ey,Cornwall:1974km,Ferrara:1992yc} 
(for $s=3/2$ see also~\cite{Deser:2000dz}), we now show that it in fact
fails for $s=2$. 

We begin by examining the massive $s=2$ propagator
\be
D^{\rm F}_{\m\n,\r\s}(p)=\frac{-i}{p^2+m^2}\,
\Big[
\Pi_{\m\r}\Pi_{\n\s}
-\frac{1}{3}\,
\Pi_{\m\n}\Pi_{\r\s}
\Big]\, ,
\qquad
\Pi_{\m\n}\equiv \eta_{\m\n}+\frac{p_\m p_\n}{m^2}
\ee
(suppressing obvious symmetrizations over $(\m\n)$ and $(\r\s)$ on the
right hand side). The terms in the 
numerator of the form 
\be
\frac{p_\m p_\r\eta_{\n\s}}{m^2}
\label{quad}
\ee
(plus permutations) must cancel in, for example, 
a tree level Compton scattering diagram since for generic kinematics they 
make unitarity-violating contributions~$\sim E^2$.
However, at a vertex, they amount to a gauge transformation, 
\be
\delta \phi_{\m\n}=\d_{(\m}\xi_{\n)}\, .
\label{Gauge}
\ee
Hence a cancellation occurs if the leading E/M vertex is 
transverse with respect to this transformation.
For spins~1 and~3/2, invariance (up to soft terms $\sim m$ or $\sim m^2$)
with respect to a single derivative
gauge invariance is sufficient for tree unitarity. However for $s\geq2$,
propagators involve increasing powers of $m^{-1}$. In particular for
$s=2$, quartic terms
\be
\frac{p_\m p_\n p_\r p_\s}{m^4}
\label{erkelsnup}
\ee
in the propagator numerator must cancel for tree unitarity.
Therefore it is necessary to also require invariance 
of the vertex under a double derivative
gauge transformation 
\be
\delta \phi_{\m\n}=\d_{(\m}\d_{\n)}\,\xi\,  .
\label{Gauge2}
\ee
For $s=2$, invariance of the leading E/M interaction couplings up to 
soft terms proportional to $m^2$ when the on-shell conditions are
imposed for any external lines would be sufficient to ensure tree unitarity.


Studying the E/M vertex with a single on-shell $s=2$ and photon line 
is equivalent to examining the terms in the field equation linear in $A_\mu$
\be
\frac{1}{ie}\,
{\cal G}_{\m\n}\Big|_A=
2A.\d\,\phi_{\m\n}+\eta_{\m\n}(\d^\r \!A^\s)\phi_{\r\s}-2\d_{(\m}(
A.\phi_{\n)})-2gF_{\r(\m}\phi_{\n)}{}^\r\, .\label{Afield}
\ee
Here we have imposed on-shell conditions
\be
\d.\phi_\n=0=\phi\, ,\qquad \d.A=0\, ,
\ee 
and will also employ the usual
\be
(\Box-m^2)\,\phi_{\m\n}=0=\Box A_\m\, .
\ee
Furthermore we have reinstated general non-minimal couplings
corresponding to the parameter $g\neq1/2$ to see whether any value
will help tree unitarity (even at the cost of an incorrect DoF count).

When the internal line attaches to terms~\eqn{quad} in the propagator,
the single derivative gauge transformation~\eqn{Gauge} is induced
so we must require a cancellation of the divergence of~\eqn{Afield}
\be
\frac{1}{ie}\,
\d.{\cal G}_{\n}\Big|_A=(2-g)\,F^{\m\r}\d_\r\phi_{\m\n}-g\,(\d_\m F_{\r\n})
\phi^{\m\r}-m^2 A.\phi_\n\, .
\ee
The cancellation of the leading term at $g=2$ 
is that observed in~\cite{Ferrara:1992yc},
whilst the second term involving the gradient of the field strength
can be cancelled by adding further (non-power renormalizable) E/M multipole
couplings~\cite{Ferrara:1992yc}. At this point it is tempting to require $g=2$
and declare the theory tree unitary since the final term proportional to
$m^2$ is apparently soft. 

However, we must still cancel the most dangerous terms~\eqn{erkelsnup}
whose induced
double derivative gauge transformation~\eqn{Gauge2} necessitates
a vanishing double divergence of~\eqn{Afield}. 
Even for $g=2$ and ignoring gradients of $F_{\m\n}$
we find
\be
\frac{1}{ie}\,
\d.\d.{\cal G}\Big|_A=-m^2\,(\d^\m A^\n)\,\phi_{\m\n}\, .
\ee
This term is hard in the photon momentum $k_\m$ and violates tree unitarity.
Clearly this is a disease generic to all higher spins with increasing powers
of $m^{-1}$ appearing in propagators.

Finally to verify that the parameter $g$ is indeed the gyromagnetic ratio,
we look at the
photon emission amplitude computed by taking all
lines in the E/M vertex on-shell. In terms of on-shell $s=2$ polarizations
$u_{\m\n}$ and $u^*_{\m\n}$ one finds
\be
T_{fi}=ie\,g\,F_{\m\n}\,u^*{}^{\r\m}u^\n{}_\r
\ee
which is consistent with low energy theorems
(see~\cite{Deser:2000dz} for details)  
\be T_{fi}=(i\mu/2s)\,F_{\m\n}\,{\cal M}^{\m\n}\, ,\ee where the Lorentz
generators in our covariant representation are
\be{\cal M}^{\m\n}\equiv 2ms\,u^*{}^{\r[\m}u^{\n]}{}_\r\ee and the magnetic
moment of a spin $s$ particle is \be\mu=(egs/2m).\ee 
Since the above result is deduced from an overall normalization only,
we also studied the soft photon limit of the 
$s=2$ Compton scattering amplitude and found
precise agreement with low energy 
theorems~\cite{Low:1954kd,Gell-Mann:1954kc,Weinberg:1970bu} 
for the identification of
the gyromagnetic ratio quoted above. We also confirmed
the above tree unitarity failure independently 
by evaluating an explicit Compton amplitude.

To summarize, 
the DoF-preserving model
of~\cite{Federbush:1961} 
yields $g=1/2$ and no choice of gyromagnetic ratio, DoF-preserving or
not, yields a tree unitary model.
For effective phenomenological applications, tree unitarity only 
signals the scale at which the effective description breaks down.
However, {\it only} the $g=1/2$ model preserves DoF and thereby offers a
reliable effective perturbative description. In any case, even this theory
lacks full consistency due to its causality difficulties.
The phenomenogical usefulness of the pure $s=2$ model coupled
to electromagnetism therefore
seems rather limited.




\section{Conclusions} 

\label{pig}

We have reexamined E/M  and (when possible) gravitational
interactions
of massive higher spin fields.
There are two possible obstructions
to consistency. The first is a violation of the constraints leading
to unphysical ghostlike propagating degrees of freedom. For $s=3/2$
this difficulty is easily avoided, but for $s=2$ it already implies a unique
$g=1/2$ magnetic moment coupling and rules out consistent interactions
with combined Maxwell--Einstein backgrounds.
The second problem, relevant {\it only} 
when the first is absent, is a breakdown
of causality/unitarity. 

For $s=3/2$ we have presented new causality bounds in 
Maxwell--Einstein backgrounds. Allowing the electromagnetic background to 
interact with an Einstein one does improve the causality properties
of the $s=3/2$ system. Indeed, one no longer needs to place unphysical
bounds on the E/M field strength, but rather only on the parameters
$(m^2,e,\kappa,\Lambda)$
of the theory . The resulting $(m^2,\Lambda)$
causality/unitarity phase diagrams (Figures~\ref{jam2},~\ref{jam3})
are generalizations of those found 
for higher spins in constant curvature 
backgrounds~\cite{Deser:2001us,Deser:2001pe}. Nonetheless,
these results provide no solace in the search for
a consistent effective theory of massive $s=3/2$ with
realistic values of the mass $m^2$ and cosmological constant $\Lambda$.
We also noted that the improved properties of massive $s=3/2$
are probably simply indications that it is a softly enough broken
version of an underlying consistent theory: ${\cal N}=2$ cosmological
SUGRA.

For $s=2$ there is much less to be done: Maxwell--Einstein
causality analysis is not even applicable as the DoF problem strikes first!
The only E/M model remaining is the gyromagnetic ratio
$g=1/2$ one of~\cite{Federbush:1961} and
for completeness we have included a derivation of its causality failure
using the method of characteristics. There have been suggestions that
$g=2$ ought to yield preferable behaviour. Certainly, this is plausible 
for an ultimately consistent description of relativistic charged higher spins
because: (i) Optical and  low energy theorems imply 
$g=2$~\cite{Weinberg:1970bu}. 
(ii) Massive higher spin string states couple to E/M backgrounds with 
$g=2$~\cite{Ferrara:1992yc}.
(iii) Fundamental $s\leq1$ particle excitations observed to date couple 
with $g=2$. We are not proposing  the pure charged $s=2$ system
as a counterexample to the $g=2$ folklore, but rather reiterate 
that it is simply not a consistent theory. Our finding that no gyromagnetic
ratio yields tree unitary amplitudes is hardly surprising in this light.
Instead, a whole Regge trajectory, as in string theory, 
is likely to be necessary for tree unitarity to hold~\cite{Cucchieri:1995tx}.
In that context, it is reasonable to speculate that $g=2$ is germane to
all spins~\cite{Ferrara:1992yc}.

An obvious open problem is to show that no finite tower of 
massive local higher spins can couple consistently to electromagnetism.
A simpler version of this problem would be to demonstrate explicitly
that spins~5/2 or~3 cannot be coupled 
to E/M backgrounds in a DoF preserving way.
Preliminary investigations indicate that this is probably the case; there
is simply no non-minimal coupling available that can restore the consistency
of the ever increasing set of constraints required for higher spins.
For example, massive $s=3$ depends on constraints 
built from single, double and triple
divergences of the field equations and the latter two are liable to become
a propagating field equations for unphysical DoF when interactions are added.
Clearly this difficulty is only compounded when $s$ increases.

\section{Acknowledgements}
It is a pleasure to thank Bernard de Wit and
especially
Massimo Porrati for discussions in the early stages. This work was
supported by the National Science Foundation under grant PHY99-73935.

\begin{appendix}

\section*{Appendix: E/M Lichnerowicz Wave Operators}

\label{zunge}

It is well known that in constant curvature backgrounds, the algebra
of covariant derivatives $D_\m$ and the Laplacian $D^2$ is vastly
simplified by introducing the wave operators of 
Lichnerowicz~\cite{Lichnerowicz:1961}. The same construction can be 
generalized to the case of electromagnetic backgrounds:
Introduce the operators $\DL{n}$ acting on symmetric $n$-index tensors
$\phi_{\m_1\ldots\m_n}$ as
\be
\DL{n}\,\phi_{\m_1\ldots\m_n}\equiv
D^2 \,\phi_{\m_1\ldots\m_n}+
2ie\,n\, F_{(\m_1}{}^\s\,\phi_{\m_2\ldots\m_n)\s}\, .
\label{lick}
\ee
It is not difficult to show that 
the following properties hold
\bea
\eta^{\m_1\m_2}\,\DL{n}\,\phi_{\m_1\ldots\m_n}&=&
\DL{n-2}\,\phi^\m{}_{\m\m_3\ldots\m_n}\, ,\qquad
(n\geq2)
\\
\DL{n}\,\eta_{(\m_1\m_2}\,\phi_{\m_3\ldots\m_n)}&=&
\eta_{(\m_1\m_2}\DL{n-2}\,\phi_{\m_3\ldots\m_n)}\, ,
\\
D^{\m_1}\,\DL{n}\,\phi_{\m_1\ldots\m_n}&=&
\DL{n-1}\,D.\phi_{\m_2\ldots\m_n}\nn\\
&+&ie\,j.\phi_{\m_2\ldots\m_n}
-2ie\,(n-1)\,(\d_\r F_{\s(\m_2})\phi_{\m_3\ldots\m_n)}{}^{\r\s}
\, ,\quad
(n\geq1)\nn\\
\\
\DL{n}\,D_{(\m_1}\,\phi_{\m_2\ldots\m_n)}&=&
D_{(\m_1}\DL{n-1}\,\phi_{\m_2\ldots\m_n)}+ie\,j_{(\m_1}
\phi_{\m_2\ldots\m_n)}\, ,
\eea
where $j_\m\equiv \d^\r F_{\r\m}$ is the E/M current.
These identities simplify further for constant
E/M field strengths.

\end{appendix}

\bibliographystyle{h-elsevier2}
\bibliography{causal}

\end{document}